\documentclass[journal=langd5,manuscript=article]{achemso}

\usepackage{amsmath,amssymb}
\usepackage{color}

\newcommand*{\dd}{\mathrm d}
\newcommand*{\ww}{\mathrm w}
\newcommand*{\hh}{\mathrm h}

\newcommand*{\sGL}{\sigma_{\scriptscriptstyle\rm GL}}
\newcommand*{\sSL}{\sigma_{\scriptscriptstyle\rm SL}}
\newcommand*{\sSG}{\sigma_{\scriptscriptstyle\rm SG}}
\newcommand*{\Scap}{S_{\scriptscriptstyle\rm CAP}}

\newcommand*{\ttc}{\theta_{\scriptscriptstyle\rm C}}

\author{Heitor C. M. Fernandes}
\author{Mendeli H. Vainstein}
\author{Carolina Brito}
\email{carolina.brito@ufrgs.br}
\affiliation[Universidade Federal do Rio Grande do Sul]
{Instituto de F{\'\i}sica, Universidade Federal do Rio Grande do Sul CP 15051, 91501-970 Porto Alegre RS, Brazil}

\title[On the Modeling of Droplet Evaporation]
  {On the Modeling of Droplet Evaporation on Superhydrophobic Surfaces}

\begin{document}






\begin{abstract}
  When a drop of water is placed on a rough surface, there are two possible extreme regimes of wetting: the one called Cassie-Baxter (CB) with air pockets trapped underneath the droplet and the one characterized by the homogeneous wetting of the surface,  called the Wenzel (W) state. A way to investigate the transition between these two states is  by means of evaporation experiments, in which the droplet starts in a CB state and, as its volume decreases, penetrates the surface's grooves, reaching a W state. Here we present a  theoretical model based on the global interfacial energies for  CB and W states that allows us to predict the thermodynamic wetting state  of the droplet for a given volume and surface texture.   We first analyze the influence of the surface geometric parameters on the droplet's final wetting state with constant volume, and show that it  depends strongly on the surface texture. We then vary the volume of the droplet keeping fixed the geometric surface parameters to mimic evaporation and show that the drop experiences a transition from the CB to the W state when its volume reduces, as observed in experiments. To investigate the dependency of the wetting state on the initial state of the droplet, we implement a cellular Potts model in three dimensions. Simulations show a very good agreement with theory when the initial state is  W, but it disagrees when the droplet is initialized in a CB state, in accordance with previous observations which show that the CB state is metastable in many cases. Both simulations and theoretical model can be  modified to study other types of surface.
\end{abstract}

\section{Introduction}

When characterizing a superhydrophobic surface, two criteria are important:  the contact angle of a drop of  liquid deposited on it should be large (typically $> 150^{\circ}$) and the hysteresis effect should be small, to ensure the drop's mobility.  One way to develop surfaces with these two properties is to texturize them with micro(nano)-patterns; with properly chosen texturing, hydrophobic surfaces can became superhydrophobic.~\cite{Liu2014}
 In many experiments in which a droplet is placed on such a substrate, it is generally found in one of two states: the Cassie-Baxter state (CB)~\cite{Cassie1944},  where it is suspended and does not come into contact with the surface's bottom, trapping air inside the surface grooves,  and the Wenzel state (W)~\cite{Wenzel1936}, characterized by the homogeneous wetting of  the surface. Transition between these two states has been observed in experiments when the droplet evaporates~\cite{McHale2005,Nosonovsky2007,Tsai2010, Xu2012}, beginning in the CB state, penetrating the surface's grooves and reaching a W state as its volume decreases. Both extreme wetting states have interesting applications,  as for example the W state is important for inkjet printing~\cite{Park2006} and coating 
 while the CB state is desirable for self-cleaning purposes~\cite{Blossey2003}, anti-freezing~\cite{Meuler2010} and production of selective condensation surfaces.~\cite{Varanasi2009}

Many experimental works~\cite{McHale2005, Nosonovsky2007,Liu2014,Tsai2010,Weibel2010, Xu2012, Li2007, Quere2008, Ramos2009, Ramos2015}, theoretical models and simulations~\cite{deGennes1985,Patankar2003,Marmur2003,Kong2006,Kusumaatmaja2007,Lundgren2007,Koishi2009,Koishi2011,Shahraz2012,Giacomello2012,Shahraz2013,Lopes2013,Mortazavi2013,Shahraz2014}  have been proposed to improve understanding of the ingredients necessary to produce superhydrophobic textured surfaces. Recently, attention has been given to the design of surfaces which minimize the contact time of an impacting droplet, in order to improve the surface's anti-icing and self-cleaning properties.~\cite{Bird2013,Liu2014a}
Experiments have been performed to understand and control the evaporation dynamics  of water droplets~\cite{McHale2005, Ramos2015}, as well as the condensation of water on superhydrophobic surfaces~\cite{Nosonovsky2007,Enright2012}.   Less attention has been given to simulations in  three dimensions in which lattice Boltzmann~\cite{Dupuis2005,Sbragaglia2007} methods are feasible but molecular dynamics simulations (MD) are costly. Several theoretical and simulation approaches use 2D~\cite{Oliveira2011, Lopes2013,Mortazavi2013}, or quasi-2D~\cite{Shahraz2013},  systems to simplify the calculations or speed up simulations. For some situations, these approaches provide results in good agreement with experimental observations; however, the use of a 2D model of a three dimensional system can mask important aspects of the real system. For example, in 2D systems, the wetting of grooves consists of independent events, whereas in three dimensions the grooves are connected and wetting is facilitated when compared to the 2D case.

In this work, we develop a theoretical continuous model that takes into account all the interfacial energies associated to the CB and W states of a water droplet deposited on microtextured surfaces. By minimizing energies associated with W and CB states, subject to the pinning of the contact line, the wetting state that globally minimizes the energy is obtained. This simple thermodynamic approach does not take into account the fact that the final state of the droplet depends on its initial condition~\cite{Quere2008, Koishi2009} as observed experimentally. To address this issue we implement 
  Monte Carlo (MC) cellular Potts model simulations~\cite{Graner1992, Glazier1993}  of a droplet in three dimensions for comparison. We first apply this model for a fixed  volume of the droplet and find a transition between the  CB and the  W state when the surface's geometrical parameters change. We then keep the microstructured surface fixed and vary the initial volume of the droplet  to mimic an evaporation experiment, finding a transition from the CB to the W state as the droplet's volume decreases.


\section{The continuous model}
\label{model}

\begin{figure}[t!]
\centering
\includegraphics[width = 0.45\columnwidth]{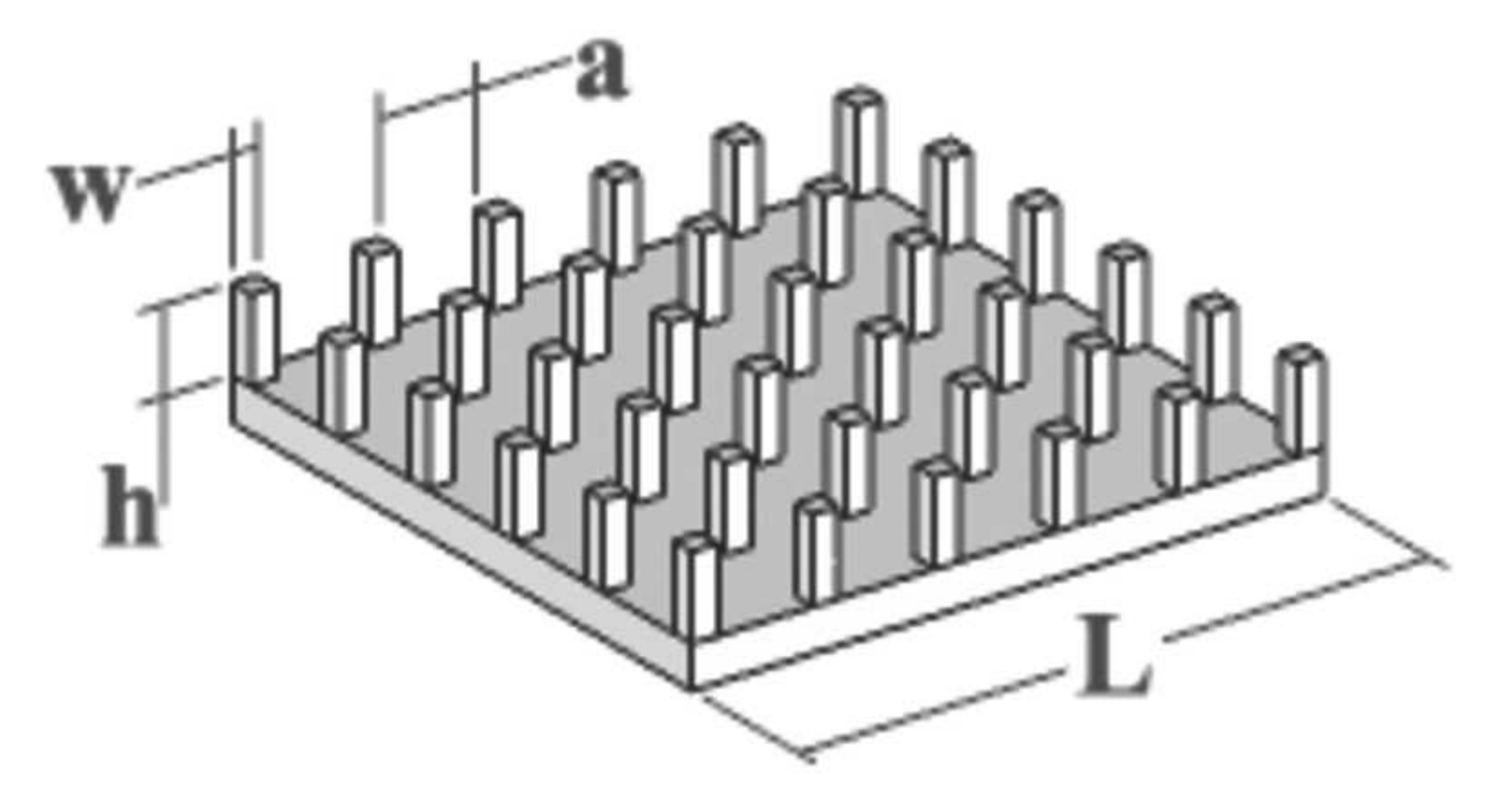}
\includegraphics[width = 0.45\columnwidth]{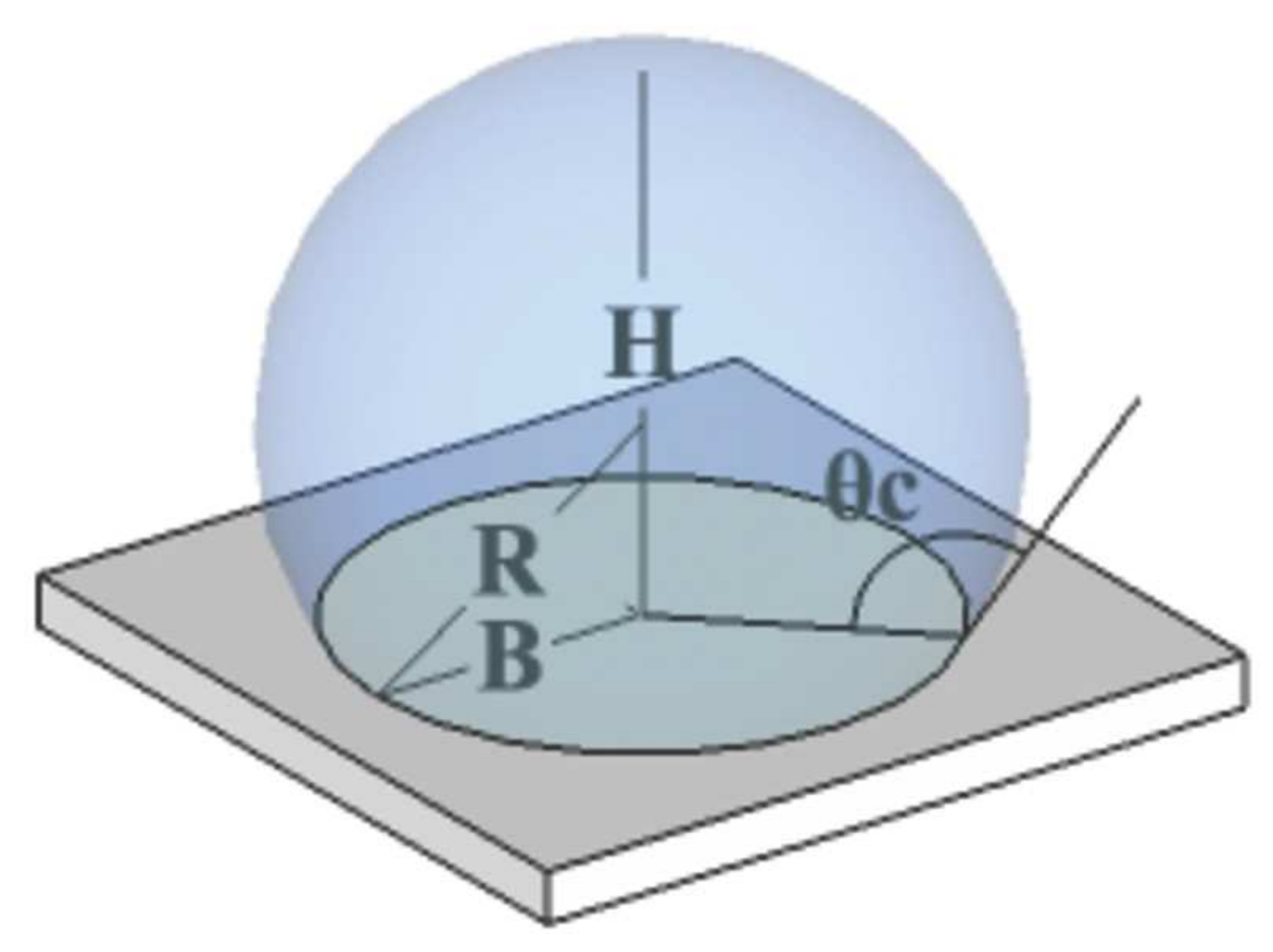}
\caption{Definition of the geometric parameters. On the left: parameters
that describe the textured surface; interpillar distance, pillar width and pillar height are represented by $a$, $w$ and $h$, respectively.  On the right:  geometric parameters of the droplet. We consider that the droplet assumes the shape of a spherical cap with radius $R$, base radius $B$, height $H$ and contact angle $\theta_C$.
}
\label{geometry}
\end{figure}

In this section we develop a model that takes into account all the energies related to the presence of interfaces when a droplet is placed on a pillared surface. 
To decide which wetting state (W or CB) is favorable from the thermodynamic point of view, we perform global energy minimization. 
Similar ideas were used in previous works to explain equilibrium properties and establish a criterion for a liquid film to propagate on a solid~\cite{Quere2008}, to determine the transition state in the specific case of each surface~\cite{Sbragaglia2007, Koishi2009, Tsai2010} and to explore the phase diagram of a two-dimensional surface with varying geometrical parameters~\cite{Shahraz2012}.  Here we consider a three dimensional droplet on a microtextured surface as the one shown in Figure~\ref{geometry}. 
The total energy of each state is given by the sum of all energies involved in creating interfaces between the droplet and  the surface on which it is placed. The difference in  energy of the system with and without the droplet on the surface can be written as 
\begin{equation}
\Delta E^{\rm s } = \Delta E_{\bf int}^{\rm s} + E_{\bf g}^{\rm s},
\label{totE}
\end{equation}
where the superscript $s$ represents the state  ($s$=W or $s$=CB), $E_{\bf g}$  is the gravitational energy and $\Delta E_{\bf int}$ is the difference in the interfacial energy between every pair formed from  liquid, solid, and  gas after the droplet is placed on the surface in  state $s$ and the energy of the surface without the droplet.  The  energy terms in eq~\ref{totE} for  the different wetting states  are given by:

\begin{eqnarray}
\Delta E_{\bf int}^{\bf {CB}} &=& N^{\rm CB} ~[~ \ww ^2 (\sSL - \sSG) +  (\dd^2-\ww^2)\sGL~] +  \sGL \Scap ^{\rm CB},\label{en_CB}\\
\Delta E_{\bf int}^{\bf {W}} &=& {\rm N}^{\rm W}  ( \dd ^2 + 4 \ww \hh )(\sSL - \sSG)  +   \sGL \Scap^{\rm W},\label{en_W}\\
E_{{\bf g}, {\scriptscriptstyle\rm CAP}}^{\rm s} &=& -\frac{\pi}{12} \rho {\rm g}  {R^{\rm s}}^3  \left \lbrace [\hh - R^{\rm s}~\cos(\ttc^{\rm s})]~ [9\cos(\ttc^{\rm s}) -\cos(3\ttc^{\rm s})-8] - 3 R^{\rm s}\sin^4(\ttc^{\rm s})\right\rbrace ,\label{en_g_cap} \\
E_{\bf g}^{\bf{CB}} &=& E_{{\bf g},{\scriptscriptstyle\rm CAP}}^{\rm CB}  \label{en_g}\\
E_{\bf g}^{\bf{W}} &=& E_{{\bf g},{\scriptscriptstyle\rm CAP}}^{\rm W} + \frac{1}{2} {\rm N}^{\rm W} \rho {\rm g}  \hh^2  ( \dd ^2 - \ww ^2 ), \label{en_g_W} 
\end{eqnarray}
where $\rho$ is the density of water,  $g$ is the acceleration of gravity,  $d= w + a$, and $\sSL$, $\sSG$,  $\sGL$ are the interfacial tensions  for the liquid-solid, solid-gas, and liquid-gas interfaces, respectively. The total number of pillars underneath the droplet is
 $ N^{\rm s} = \frac{\pi}{4}(2 B ^{\rm s}/d)^2$, where 
$B^{\rm s} = R^{\rm s}\sin(\ttc^{\rm s})$  is the base radius. The surface area of the spherical cap in contact with air is given by 
$\Scap^{\rm s}=2\pi  {R^{\rm s}}^2 [1-\cos (\ttc^{\rm s})]$. We emphasize that the gravitational energy $E_{\bf g}$ is much smaller than the interfacial energies.  Actually, when the droplet is on the surface, it is negligible in the cases under study; it is only explicitly presented for completeness and because in the simulation 
 it is important when the droplet is not in contact with the surface.

\subsection{ Energy minimization ~~}
We now use this model to address the following question: when a droplet with a fixed volume $V_0$ is placed on a surface as the one  shown in  Figure~\ref{geometry}, what should be its final wetting state, W or CB? Note that, for a surface with a given geometry, the parameters $h$, $a$ and $w$ are fixed and its chemical properties define the interfacial tensions. Therefore, given these values, the energies only depend on the droplet radius $R^{\rm s}$ and contact angle  $\ttc^{\rm s}$. For a given droplet volume, the thermodynamically stable wetting state is the one with the lowest energy.  
 
For each surface and initial volume $V_0$, we first compute the energy that the droplet would have in the CB state. To do so, we minimize 
$\Delta E^{\rm CB}$ in respect to $R^{\rm CB}$ and $\ttc^{\rm CB}$. 
The procedure is the following: we consider an initial volume $V_0$ and then vary the contact angle $\ttc^{\rm {CB}} \in (0, \pi]$. For each 
contact angle $\ttc^{\rm {CB}}$ and volume $V_0$, we compute the droplet radius $R^{\rm {CB}}$ and the energy difference  $\Delta E^{\rm {CB} }$ 
associated with these parameters using  eqs~(\ref{en_CB}), (\ref{en_g_cap}) and (\ref{en_g}). To decide which is the minimum energy for the CB state  
$\Delta E^{\rm {CB} }_{min}$, we also impose the  constraint that the contact line of the droplet has to be pinned to the pillars~\cite{Shahraz2012}. 
 This implies that the base radius $B^{\rm {CB}}$ does not have a continuous value as a function of volume.
 The procedure to determine $R^{W}$ and $\Delta E^{\rm W}_{min}$ in the minimum energy W state is similar: in this case, we use eqs~(\ref{en_W}), (\ref{en_g_cap}) and (\ref{en_g_W}),
 which includes, in addition to the spherical cap, the volume of water present below 
 it in the interpillar space. Mathematically, the difference amounts to a solution of a complete cubic equation in $R^W$ for the W state as opposed to the solution of a 
 simpler cubic equation of the form $R^{\rm CB}=\{ 12 V_0/ [ \pi( \cos(3\ttc^{\rm CB}) -9 \cos(\ttc^{\rm CB}) + 8 ) ]  \}^{1/3} $ for the CB state. The thermodynamically stable state is then the one with the lowest $\Delta E_{min}$.
 Once it is defined, all geometric parameters of the droplet (contact angle $\theta_C$, radius $R$, base radius $B$, spherical cap height $H$) are determined and can be compared with those obtained from numerical  simulations.

\section{Simulations: The Cellular Potts  model}
\label{simus}

Many different approaches have been considered in the simulation of water droplets on a hydrophobic surface, such as  molecular dynamics (MD)~\cite{Koishi2009,Wu2009,Shahraz2012,Shahraz2013}, lattice Boltzmann methods~\cite{Dupuis2005,Sbragaglia2007}, and the finite-element method for the solution of the Navier-Stokes equations~\cite{Wind-Willassen2014}. Recently, Monte Carlo (MC) simulations of the cellular Potts model (CPM)~\cite{Graner1992,Glazier1993} have been used to model droplets  on superhydrophobic surfaces~\cite{Oliveira2011,Lopes2013,Mortazavi2013}. 
The advantage of MC simulations is that, since atoms are not explicitly simulated as opposed to MD simulations,  it is a more consistent framework to treat mesoscopic systems by allowing a coarse-grained approach. This regime is more appropriate for comparison with experimental results. In this spirit, Ising and Potts models have been used to simulate complex systems ranging from the simulation of biological cells~\cite{Graner1992} to the complex behavior of leaky faucets~\cite{Oliveira1993}, among many other applications such as capillary evaporation~\cite{Luzar2000}, study of growth regimes of two-dimensional coarsening foams~\cite{Fortuna2012} and solution of advection-diffusion equations~\cite{Dan2005}.

We follow a previous 2D cellular Potts model~\cite{Oliveira2011,Lopes2013,Mortazavi2013} and we further extend it by considering its 3D counterpart to model consequences of characteristics not present in 2D systems, such as groove connectivity. With this in mind, extrapolation of results in simulations done in 2D to systems in 3D is not straightforward and should be made with care. For example, in 2D, a transition from a CB state to a W one requires the wetting of independent grooves beneath the droplet, a situation that can be compared to the wetting of nanoporous alumina in which the pores are independent from one another~\cite{Ran2008,Weibel2010}. In 3D patterned structures, once a groove has become wet, the liquid can propagate to other grooves; therefore, simulations in 2D can lead one to believe that energy barriers are greater than they really are.

We model our system by a three state cellular Potts model on a simple cubic lattice 
 in which each state represents one of the components: liquid, gas and solid.  This simple approach presents the basic features necessary to describe the physical situation, \emph{i.e.}, the existence of interfaces between gas/liquid/solid in the presence of the gravitational field.
The Hamiltonian used is
\begin{equation}
H = \frac{1}{2} \sum_{<i,j>} E_{s_i,s_j} (1-\delta_{s_i,s_j}) + \lambda (\sum_i \delta_{s_i,1} - V_T)^2 + mg \sum_{i} h_i ~\delta_{s_i,1}.
\label{eq.potts}
\end{equation}

Here, the spins $s_i \in \{0,1,2\}$ represent gas, liquid and solid states, respectively. The first summation ranges over pairs of 
 neighbors which comprise the 3D Moore neighborhood in the simple cubic lattice ($26$ sites, excluding the central one); $E_{s_i,s_j}$ are the interaction energies of $s_i$ and $s_j$ of different states at interfaces and $\delta_{s_i,s_j}$ is the Kronecker delta.  $V_T$ stands for the target volume of the droplet  and $\lambda$ mimics its compressibility; the second summation represents the total liquid volume, the total number of sites $i$ for which $s_i=1$. The last term is the gravitational energy,  where $g=10\,m/s^2$ is the acceleration of gravity.  In both the volumetric and gravitational terms, only sites with liquid, $s_i=1$, contribute.

The parameters for our Hamiltonian, eq~\ref{eq.potts}, were based on 
 those used in experiments with water on a Polydimethylsiloxane (PDMS) surface~\cite{Tsai2010} (surface tension of water $\sGL=70\, mN/m $ and $\sSG=25\,mN/m$, for the PDMS surface).  In our numerical simulations, these values are divided by the number of neighboring sites that contribute to the first summation in eq~\ref{eq.potts}, that is, 26 neighboring sites.
 Our length scale is such that one lattice spacing corresponds to $1\, \mu m$.
  This implies that the interaction energies $E_{s_i,s_j}=\sigma_{ij} A$,  with $A=1\mu m^2$ being the unit area, are given by  $E_{0,1} =2.70 \times 10^{-9}\, \mu J$,
  $E_{0,2} = 0.96 \times 10^{-9} \,\mu J$ and  
  $E_{1,2} = 1.93 \times 10^{-9} \,\mu J$. 
  The third value is obtained from Young's relation $\sGL \cos(\theta)= \sSG- \sSL$, where $\theta= 111^\circ$ is the contact angle on a smooth surface. The mass existent in a unit cube  is $m=10^{-15}kg$ and we fix 
  $\lambda = 10^{-9}\, \mu J / (\mu m)^6$.

The standard Metropolis algorithm is used to accept trial spin flips.
We keep a list of sites located at the droplet's frontier, \emph{i.e.}, at the
liquid-gas or liquid-solid  interfaces, since those sites are the only ones
that contribute to energy variations.\bibnote{The interface sites are defined as the ones which have at least one neighbor in a different state than its own, among the $18$ first and second closest neighbors.   This smaller neighborhood (compared to the Moore neighborhood) is used so that it is possible to obtain spherical droplets, despite the underlying lattice symmetries.}
For the dynamics, one site at  the interface, either liquid or gas,
is chosen at random and a change in state between liquid and gas is
accepted with probability $\min\{1,\exp(-\beta\Delta H)\}$, where $\beta = 1/T$ is the inverse of the effective temperature of the CPM~\cite{Glazier1993}, which acts as a noise to allow the phase space to be explored. A value of $T=13$ is used throughout the paper because it allows the system to fluctuate with an acceptance rate of, typically, approximately $22\%$. Each Monte Carlo step (MCS) is comprised of a number of trial spin flips equal to the total target volume of liquid, $V_T$. 
 The total run of a simulation is $10^6$ MCS, from which the last $2\times10^5$ MCS are used to measure observables of interest. It should be noted that even with this long transient time,  for some initial conditions the system does not reach the thermodynamically stable state and becomes trapped in a metastable state (see next section). 
  At least $10$ different samples evolving with  different random number sequences 
 are used for each set of simulation parameters  and for each initial condition. The presented results are averages over these distinct realizations of the simulation.

\begin{figure}[t!]
\centering
\includegraphics[width = 1.0\columnwidth]{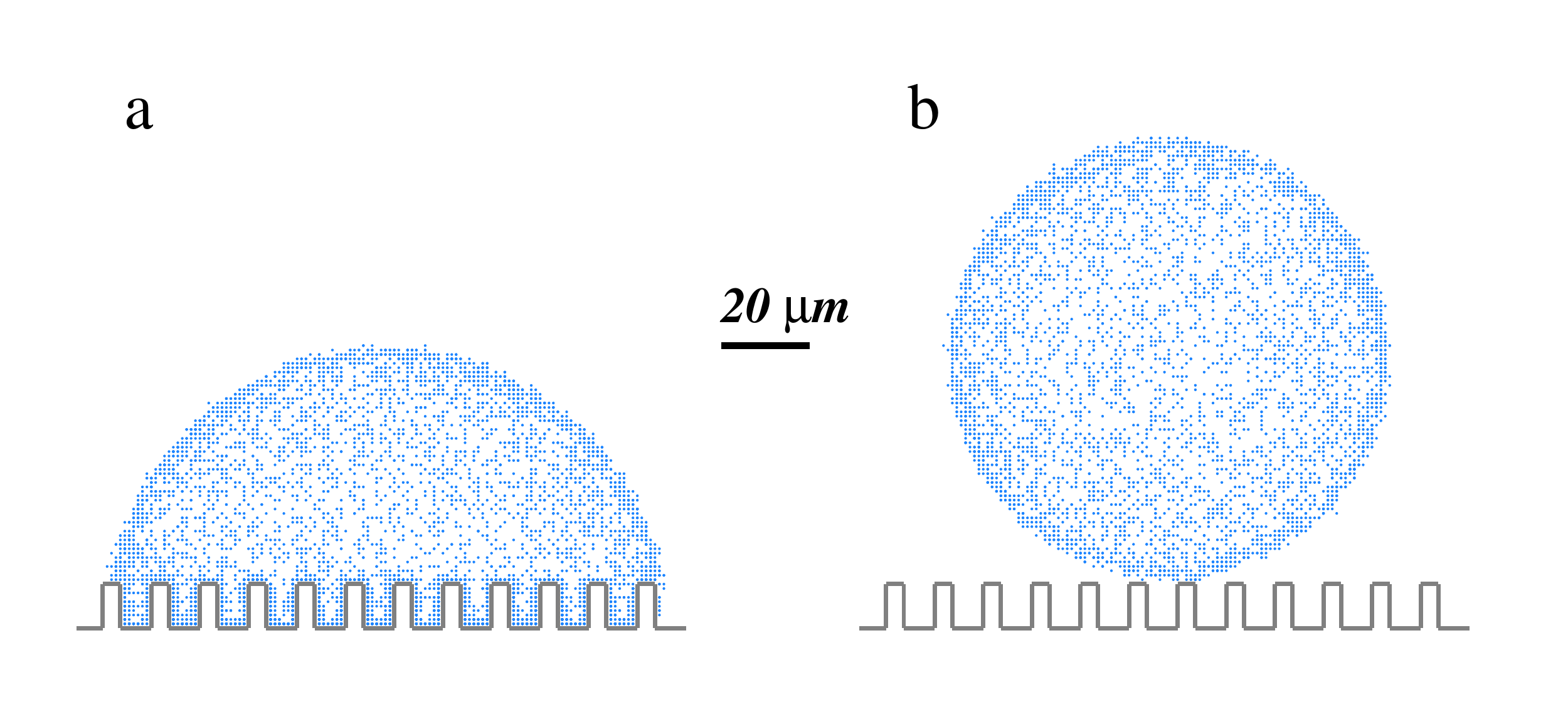}
\caption{Initial states of the Monte Carlo simulations for the (a) Wenzel and (b) Cassie-Baxter states.  For clarity, only $10\%$ of the droplet's surface sites with $R_0=50\,\mu m$ are plotted in a side view of the 3D initial configuration. Here, the height of the pillars and interpillar distance are given by $h=10\,\mu m$ and $a=6\,\mu m$, respectively. The pillar width is fixed throughout the paper at $w=5\,\mu m$.}
\label{init_config}
\end{figure}

The system is initialized in either one of the CB or W states as is exemplified in Figure~\ref{init_config}. In the former, the droplet is placed slightly above the surface and allowed to relax under the influence of gravity. In this case, the gravitational energy is important until the droplet reaches the surface. The latter state is created using a hemisphere with the same initial volume $V_0$ of the CB state, \emph{i.e.}, $V_0 \approx V_T =4/3\,\pi R_0^3$, due to the discreteness of the lattice. 
The value of $\lambda$  is chosen such that the energy fluctuations due to volume variations around $V_T$ 
 are smaller than energy changes due to spin flips at the interface. 
 In simulations with $\lambda = 10^{-9} \mu J / (\mu m)^6$, 
 fluctuations  of less than $1\%$  in volume are observed. Moreover,   with our choice of parameters, it is observed that volume changes are only pronounced in the early stages of simulations  and stabilize after, typically, $5 \times 10^4$ MCS. 

In the simulations, two basic quantities are measured: base radius and droplet height, $B$ and $H$, respectively, as shown in Figure~\ref{geometry}. These quantities are used to calculate the contact angle $\theta_C$, the number of pillars below the droplet and the drop radius $R$. It is assumed in the calculation that the droplet is a spherical cap and numerical results corroborate this assumption. 

\section{Results and discussion}
\label{results}

In this section, we compare theoretical predictions of our continuous model  
with MC simulations for two distinct kinds of experiments. In the first part of this section, we study the  wetting state of a droplet with fixed
target volume $V_T$  on  surfaces with different textures. In the second part, we analyze the effect of target volume variation keeping the surface 
texture fixed. The idea is to mimic  experiments where the droplet evaporates~\cite{McHale2005,Tsai2010,Chen2012,Ramos2015}, and to characterize the 
dependency of the wetting mode on the droplet's volume. To this end, we performed various Monte Carlo simulations for different droplet volumes, kept fixed in each run. Since this type of simulation produces a series of configurations that tend to equilibrium and the appreciable volume fluctuations in each run occur early in the simulations  for a few Monte Carlo steps during the equilibration transient, the results of a longer simulation with decreasing target volume (allowing equilibration at each volume value) are the same as the results obtained from the concatenation of a series of equilibrium simulations at different volumes, which are the results we present. 

Before applying the Potts model to simulate the droplet on a patterned surface, we verify its prediction for a flat surface. Using the parameters  presented in the previous section, 
a contact angle of $\theta_C \approx 114^{\circ}$ is obtained in our simulation, which is in good agreement with the  experimental observation~\cite{Tsai2010} of $(113 \pm 7)^{\circ}$ and very close to the value obtained from  Young's relation, $\theta = 111^\circ$.

 \begin{figure}[t!]
\centering
\includegraphics[angle=0,width=1.0\columnwidth]{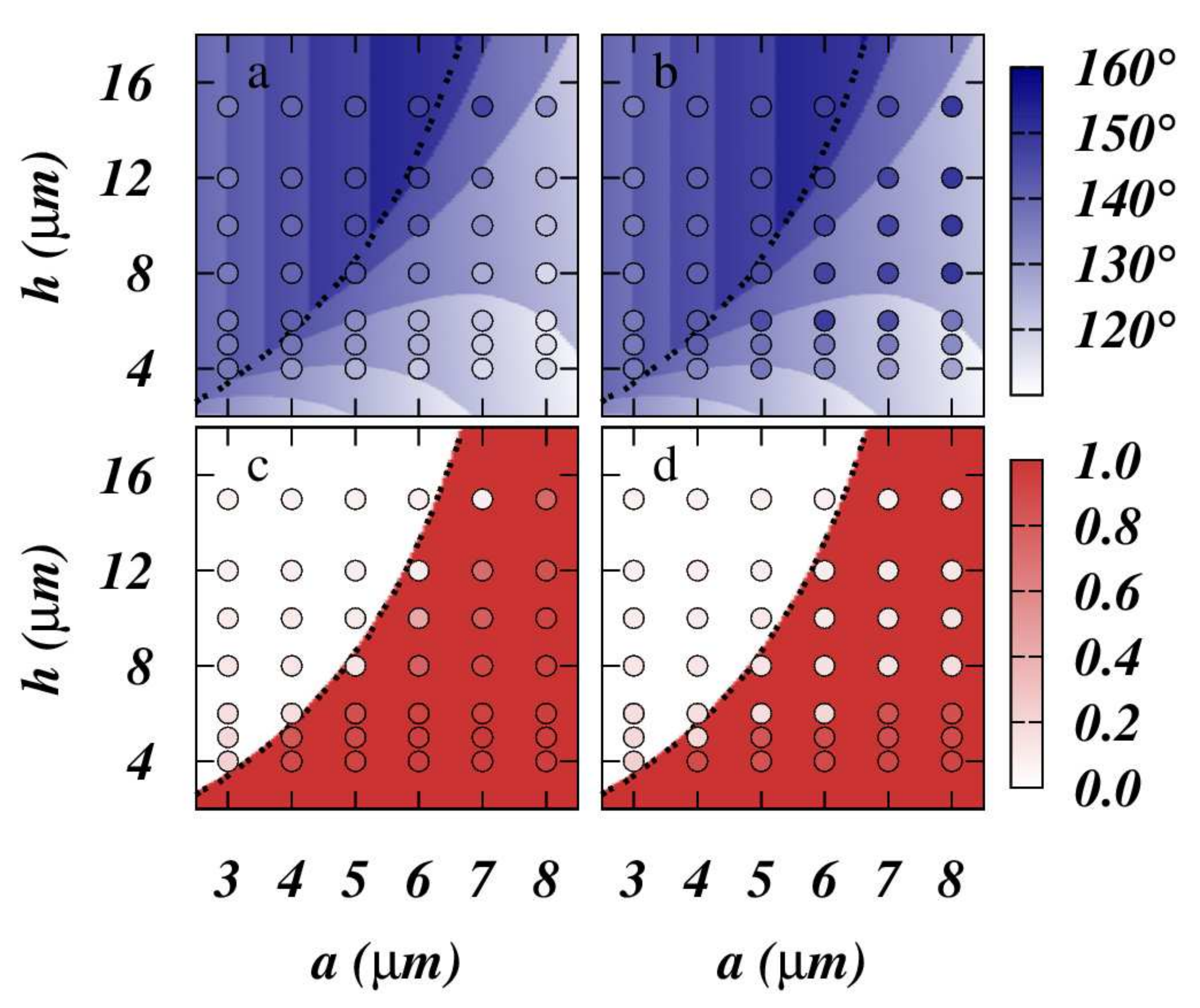}
\caption{(a) and (b): Theoretical phase diagrams of the contact angle for a droplet of initial size $R_0=50\,\mu m$ as a function of two geometrical parameters of the surface: the height of the pillars $h$ and the distance $a$ between them. 
 The dotted line shows the predicted thermodynamic transition between the Cassie-Baxter and Wenzel states, being that 
 the Wenzel state corresponds to the region below the line. The subdivisions visible inside each phase correspond to different pinning modes, (different number of pillars beneath the droplet), as previously observed~\cite{Shahraz2012}. The open symbols correspond to simulation results, averaged over $10$ runs. (a) and (c): 
  simulations start in the Wenzel configuration; 
  (b) and (d): simulations start in a Cassie-Baxter state. Along the transition line, the angle in the Cassie state is predicted to be constant.
  (c) and (d): Fraction $f$ of water below the droplet's base area.
  }  
\label{theor_diag_angle_3d}
\end{figure}

\subsection{Droplet with fixed target volume on surfaces with different textures.~~}
\label{results_volFixo}

In Figure~\ref{theor_diag_angle_3d}, we present two kinds of theoretical phase diagrams as a function of the geometric parameters of the surface. The upper diagrams represent the contact angle $\theta_C$ of the droplet  and the lower ones present the fraction $f$ of water underneath the droplet's base area, for fixed values of pillar width and initial droplet radius  ($w=5\,\mu m$ and  $R_0=50\,\mu m$, respectively)  and varying  $h \in [1,20]\,\mu m$, and $a \in [1,10]\,\mu m$. The background shades of gray (blue) present results from the minimum energy state of the continuous model,  while the circles 
   contain the results of an average over 10 different runs of the simulations. We stress that the results of the continuous model remain unchanged whether we include the gravitational energy terms or not. The contact angle $\theta_C$  diagrams (top) present subdivisions that correspond to different pinning modes, \emph{i.e.}, different numbers of pillars underneath the droplet's base area, a feature also seen in Shahraz \emph{et al.}~\cite{Shahraz2012}. 
 Points where  $E^{\rm W}_{min} = E^{\rm CB}_{min}$ define the transition line between the W and CB  states, shown as dotted lines in the diagrams.
To test the dependency of the final states on the initial conditions, we implemented simulations initializing the droplet in both the  W (left diagrams) and the CB state (right diagrams), as illustrated in Figure~\ref{init_config}. We emphasize that the theoretical diagrams (background of  Figure~\ref{theor_diag_angle_3d}) on  both sides are the same; the only difference between them are the simulation results
for each initial condition. The lower diagrams present the fraction $f$ of water beneath the droplet's base area, a quantity easily obtainable from simulations, but not usually available experimentally.  In the theoretical case, only the two extreme regimes of wettability were considered, therefore  $f$ can only assume two values, $f=1$ (W state) or $f=0$ (CB state).  However, observed deviations from the theoretical values in simulations are due to interface fluctuations.

\begin{figure}[t!]
\centering
\includegraphics[width=1\columnwidth]{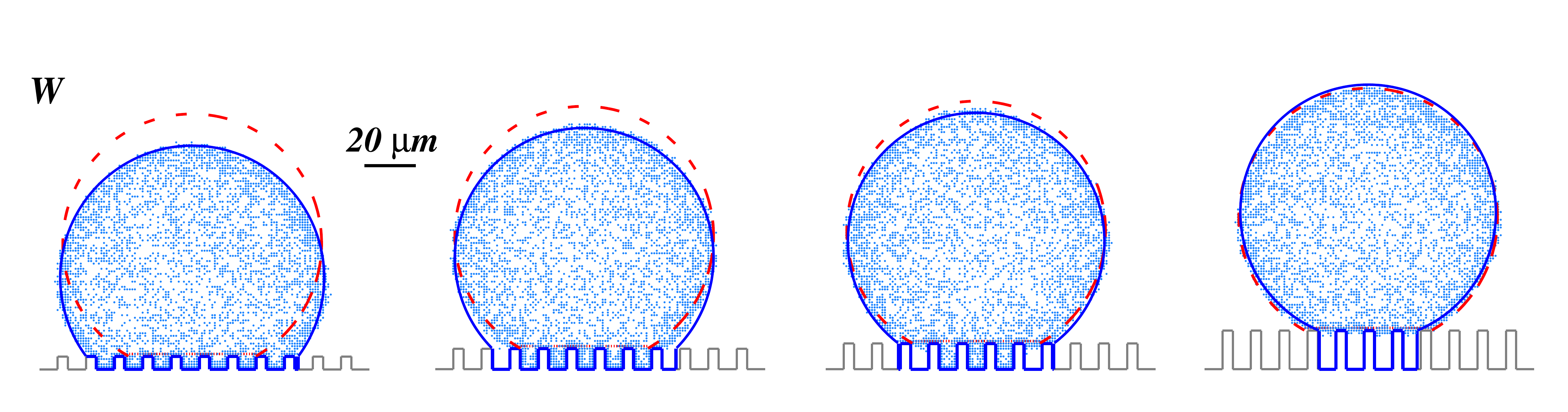}
\includegraphics[width=1\columnwidth]{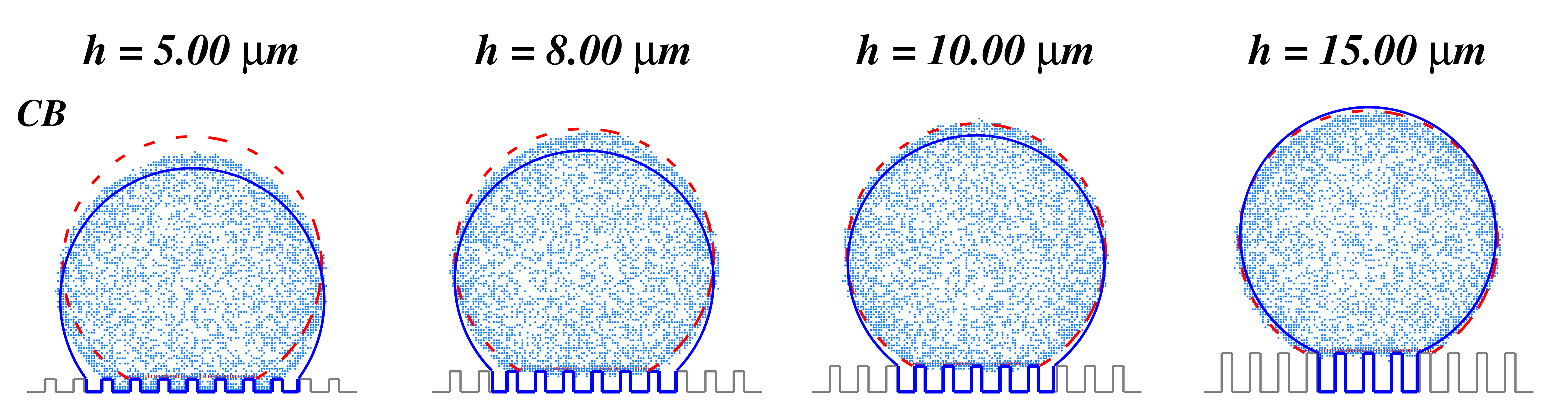}
\caption{Cross section of the droplet configuration in the final state of the Monte Carlo simulation, starting from the W configuration (above) and from the  CB configuration (below). The solid (blue) line represents the cross section for the minimum energy W configuration, while the dashed (red) line, the  one for the minimum energy CB configuration.  The snapshots correspond to droplets with $R_0=50\,\mu m$ placed on a surface with fixed interpillar distance and pillar width ($a=6\,\mu m$, $w=5\,\mu m$ ) and varying pillar height $h$. }
\label{snaps2}
\end{figure}

 All diagrams indicate a strong dependency of the droplet's wetting state on the different  surface parameters. In general lines, the CB state is less energetic than W state for small values of $a$ and large values of $h$. The same behavior was observed in  full atomistic molecular dynamics simulations~\cite{Shahraz2013, Lundgren2003}.

Let us  now compare the theoretical predictions with the results of simulations.  When the initial state is W (Figure~\ref{theor_diag_angle_3d} (a) and (c)), both quantities,  $\theta_C$ and $f$, present a good agreement with the theory. However, when the initial state is CB (Figure~\ref{theor_diag_angle_3d} (b) and (d)), the agreement between simulations and theory is very good only in the region where the thermodynamically stable state is the CB one. This means that, when the initial state is W and the thermodynamic state is CB, the droplet is able to change its state (during a simulation run, all samples reach the predicted state). On the other hand, the converse is not true; when the theoretically predicted thermodynamically stable state is W and the droplet is initialized in the CB state,  the droplet is generally not able to overcome the barrier between the states and becomes trapped in the metastable CB state, unless the pillar height is low enough.

\begin{figure}[t!]
\centering
\includegraphics[angle=0,width=1.0\columnwidth]{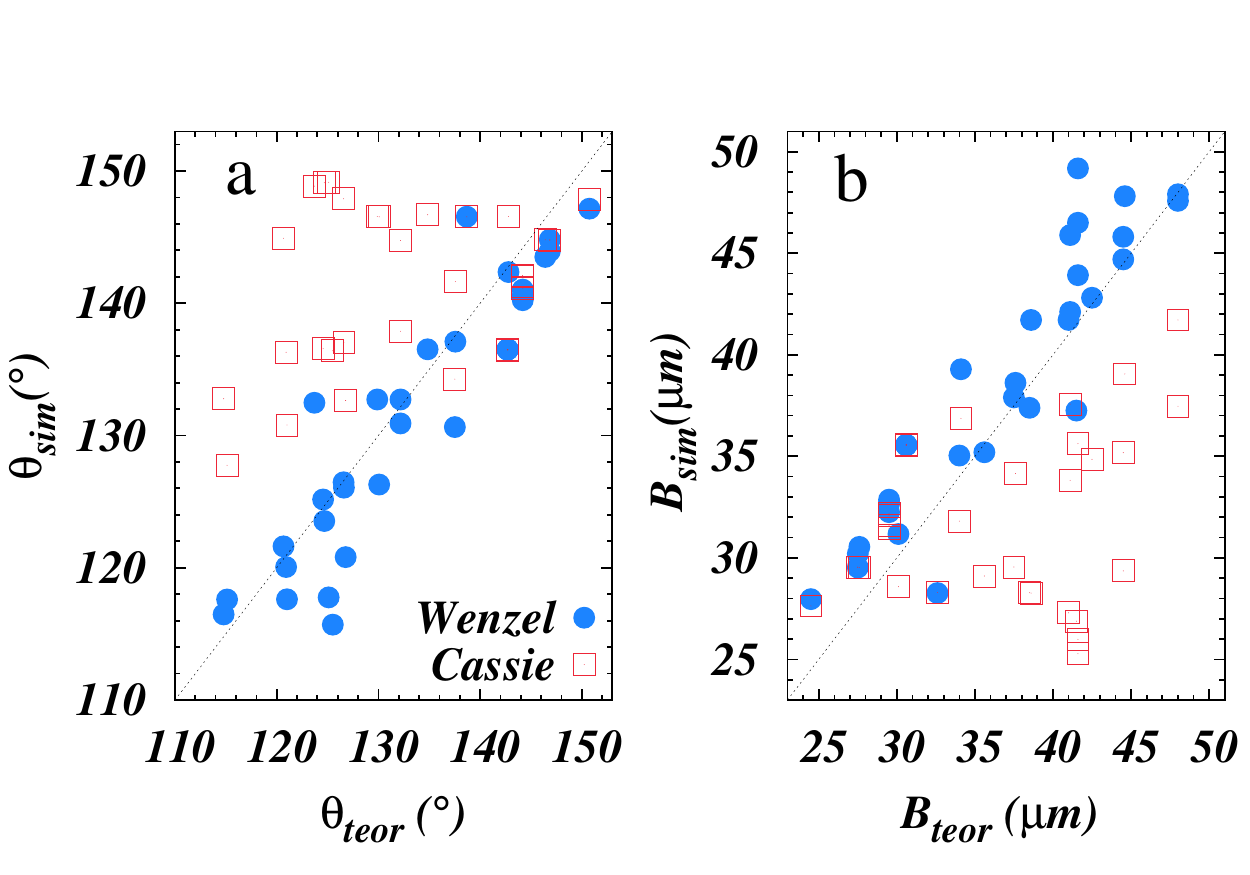}
\caption{Scatter plot of stationary (a) contact angles  and (b) base radius $B$ from simulations as a function of the theoretical values. The (blue) circles correspond to simulations starting in the W configuration while the (red) squares, to an initial CB state. The dotted line is the expected relation of  equality between simulations and theory. The simulations starting in the W state are better able to explore the phase space and to make the  transition to the CB state, thus presenting results closer to the expected values. Points are averages over $10$ simulation runs for $R_0=50\,\mu m$ and for various values of pillar height and interpillar spacing, corresponding to the points in Figure~\ref{theor_diag_angle_3d}. 
}
\label{scatter}
\end{figure}
In Figure~\ref{snaps2} we compare cross sections of final droplet configurations  for different surfaces and different initial conditions obtained from MC simulations, with the resultant cross sections  obtained from the theoretical prediction of the continuous model. 
In the snapshots we show states that correspond to  both $E^{\rm {CB} }_{min}$ (the dashed/red line) and $E^{\rm {W} }_{min}$ (continuous/blue line). The top row presents results for the case in which the initial state is W and the bottom row, for the simulations starting in the CB state. 
 The snapshots are shown for a droplet with  $R_0=50\,\mu m$  placed on a surface with fixed interpillar distance $a=6\,\mu m$ and pillar width  $w=5\,\mu m$, and varying height  $h$. The theoretical prediction is a stable W state for $h \lessapprox 13\,\mu m$ and a transition to a CB stable state thereafter. When the initial state is W, the droplet reaches the theoretically predicted state  for almost all values of $h$, except close to the transition. When the initial state is CB,  the final state of the droplet is W only for $h=5\,\mu m$.  Interestingly, for $h \ge 8\,\mu m$ the final state of the droplet  coincides with the least energetic CB state; that this should happen is not obvious, since we do not consider the possibility of other intermediate states as possible stable states in the continuous model. A quantitative display of the initial state dependence is shown in  Figure~\ref{scatter}, in which we present scatter plots of the contact angle $\theta$ and base radius $B$. The horizontal axis presents the theoretical values and the vertical one, the results from simulation averages. A good agreement is found throughout the range for the W initial state (blue circles), in contrast to the results for the CB initial state (red squares).

This observation of metastability is in agreement with the observation made in experiments~\cite{McHale2005,Quere2008} and (almost) 2D systems simulated by means of molecular dynamics of nanodroplets~\cite{Shahraz2012,Shahraz2013} and are consistent with the existence of a high energy barrier between the  thermodynamical states. As $h$ gets higher, it becomes increasingly more difficult for the system to go from the CB state to the W state. Advanced computer techniques have been  used to calculate the  free energy barrier between W and CB states in a quasi-2D nanodroplet~\cite{Shahraz2014}, which tend to be in the range from $2$  to  approximately $50\,k_{\mathrm{B}} T$. A similar effect of coexistence of  wetting states  depending on the droplet's history has been  described experimentally:  when the droplet's temperature is increased and  the system jumps from W to CB, the system becomes trapped in the metastable CB even after the  temperature is decreased back to its initial value~\cite{Liu2011}. In this spirit, our system can be interpreted as being subjected to fluctuations (controlled by the effective temperature parameter) due to an outside source of energy, which permits transitions from the W to the CB state.

\begin{figure}[t!]
\centering
\includegraphics[angle=0,width=\columnwidth]{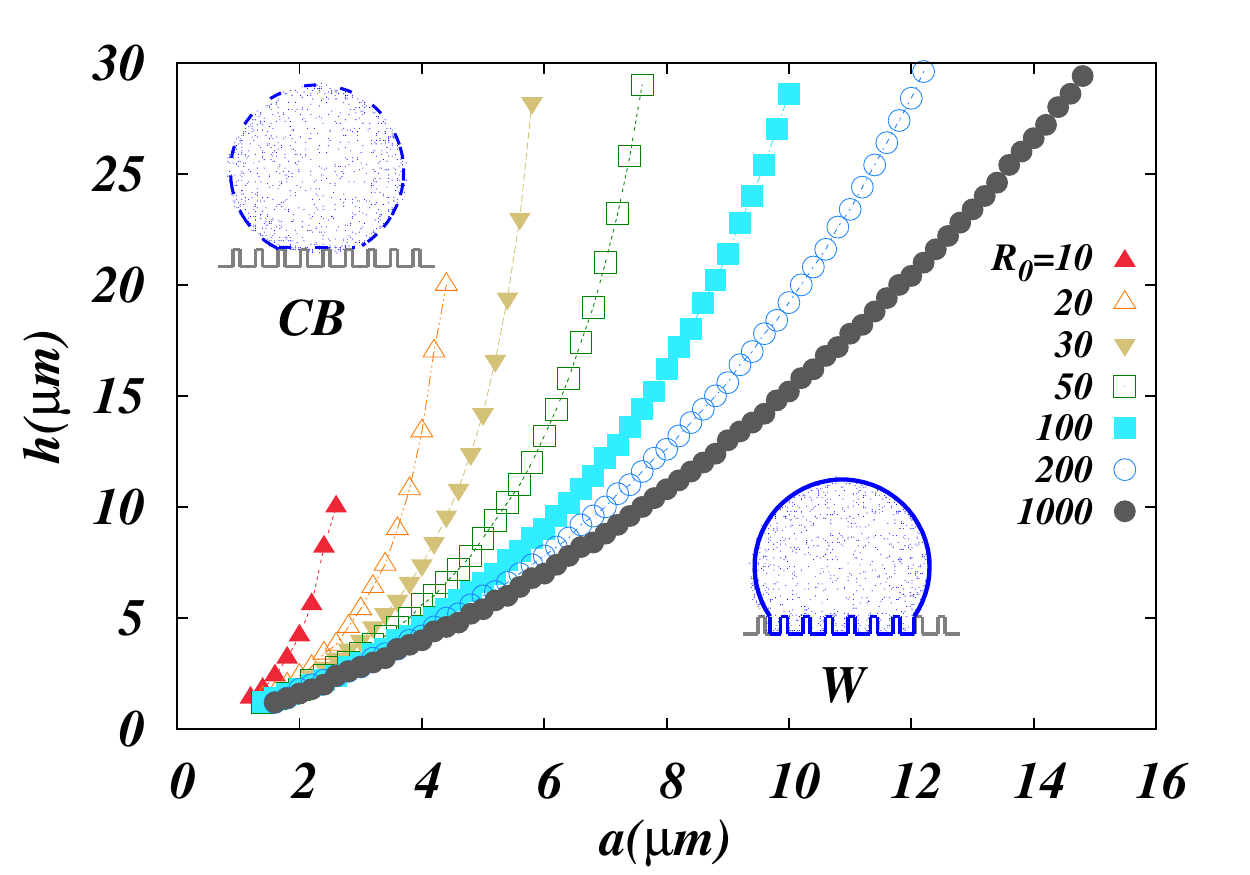}
\caption{Theoretical transition lines between CB and W states for different volumes of the  droplet. In the legend, the volume is expressed in terms of its corresponding radius $R_0$ (in $\mu m$). Note that, for increasing radius, the size of region where the CB state is stable increases. The transition lines are computed at most up to $a=R$ and $h=R$, to keep the texture of the surface comparable to the size of the droplet.}
\label{transitionR}
\end{figure}

\subsection{Droplet with varying initial volume}
\label{results_volVaria}

To take into account the role of evaporation on the droplet's final state, we note that $E^{\rm W}$, $E^{\rm {CB}}$ are modified when the droplet's volume is reduced, since the terms $N^{\rm s}$ and $S_{CAP}^{\rm s}$ in  eqs~(\ref{en_CB}) to~(\ref{en_g_W}) depend on the droplet radius $R$. During the process, the minimum energy state changes from one wetting state to the other. Figure~\ref{transitionR} presents the transition lines for several values of droplet volume as a function of the interpillar distance $a$ and the height of the pillars $h$. It shows that, the bigger the droplet, the bigger is the region the where the CB state is predicted to be the thermodynamically stable one. This observation is in agreement with experiments where the droplet changes its state from the CB to the W during its evaporation process~\cite{Tsai2010}.

\begin{figure}[t!]
\centering
\includegraphics[angle=0,width=1.0\columnwidth]{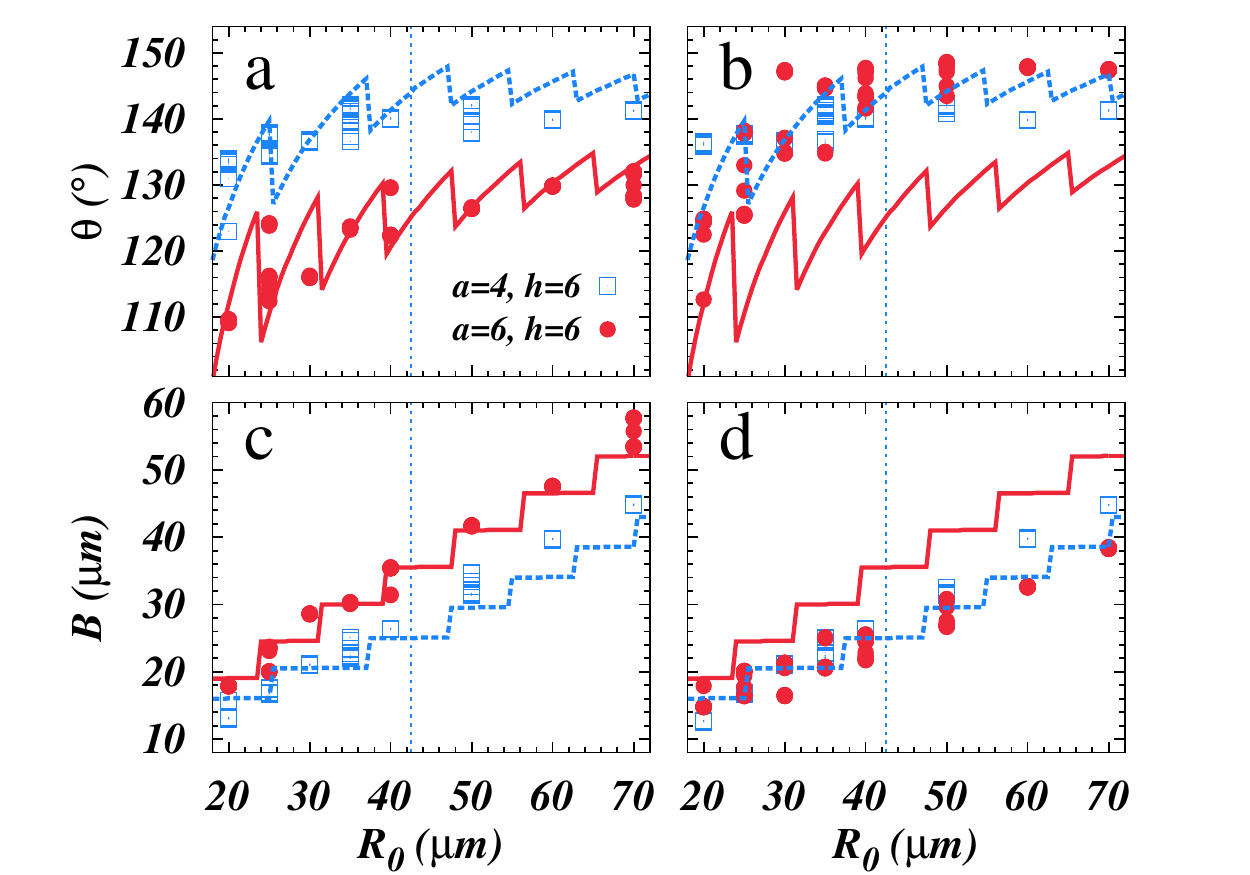}
\caption{Contact angle and base radius $B$ as a function of initial radius $R_0$. Each curve corresponds to a different structure of the underlying surface. The dashed (solid) curves correspond to the thermodynamic stable values, regardless of wetting state, obtained from the continuous model for $a=4\,\mu m,\, h=6\,\mu m$ ($a=6\,\mu m,\, h=6\,\mu m$). Points are unaveraged simulation results. The vertical dotted line represents the themodynamic transition from the Wenzel (left) to the Cassie-Baxter state (right) which occurs at $R_0\approx 42.5\,\mu m$ in the case $a=4\,\mu m,\, h=6\,\mu m$. For the case $a=6\,\mu m,\, h=6\,\mu m$, the thermodynamic stable state of the system is predicted to always be the Wenzel state for the displayed range of $R_0$, and therefore no transition line is displayed. (a) and (c): Wenzel starting configuration.  (b) and (d): Cassie-Baxter starting configuration. Note that in the case $a=6\,\mu m,\, h=6\,\mu m$, the system remains trapped close to the initial state, unable to make transition to the predicted Wenzel state, when initialized in the Cassie-Baxter configuration.   }
\label{theta_B_f_vsR}
\end{figure}

Figure~\ref{theta_B_f_vsR} shows the contact angle  $\theta_C$ and the basis radius $B$ as a function of the droplet's initial radius   $R_0$ for two different  values of the triplet $(w, a, h)$. The lines are theoretical predictions  and the symbols are the results of simulations (initial state W in (a) and (c);  CB in (b) and (d)).  We note that the stick-slip behavior of $\theta_C$ and $B$ is due to the fact that the energy is minimized  subject to the constraint that the contact line is pinned. 
 A similar behavior has been described~\cite{Ramos2015}, in which the authors study experimentally the evaporation of a droplet of water on highly hydrophobic surfaces for different temperatures and find that the droplets start evaporating in a constant contact radius mode  but then  switch to a more complex mode characterized by a series of stick-slip events.  These events are also observed  in MD simulations~\cite{Shahraz2013} and in the condensation of water on a superhydrophobic surface~\cite{Enright2012}.

\section{Conclusions}
\label{concl}

In this work we developed a simple model to understand the wetting state of a three-dimensional droplet when  placed on a microtextured surface based on the analysis of the total interfacial energies associated with the two extreme wetting states, W and CB.  We first apply this idea to study the dependency of the final state of a droplet with a fixed volume on the different textured surfaces and then 
 analyze how the state of the droplet changes when its volume reduces, mimicking evaporation. Because this thermodynamic approach does not take into account any aspect of the dynamics, we implement Monte Carlo simulations of a Potts model in three-dimensions. Previously available computer simulations are either 2D or focus on nanodroplets~\cite{Koishi2009}, with tens of thousands of particles, far from the full atomistic view of the physical situation and  allowing only a few  grooves to be located below the droplet.  The usual solution to this issue is to provide a formulation, in which a scaling of both the droplet and the underlying surface is performed at the same time~\cite{Shahraz2012,Shahraz2013}, thus rendering the analysis independent of droplet size, as a justification for the use of nanoscale molecular dynamics simulations to explain phenomena at the microscopic scale. However, this scaling approach masks the effect of system volume changes throughout simulations and does not capture some transitions from the W to the CB state, resulting in an incomplete framework to understand the effect of volume changes in evaporation experiments. The simulations  presented here allow one to have a better understanding of the role of the initial conditions on the final state of the droplet, to capture the droplet size dependent transition from the CB to the W state and to grasp some insight about the energy barriers between states. 
  
 The global thermodynamic approach has some limitations as has been observed in experiments with water condensing in superhydrophobic surfaces, which is due to the fact that the Cassie-Wenzel transition is a multiscale phenomenon governed by micro- and nanoscale effects~\cite{Nosonovsky2007,Enright2012}. Nevertheless, this approach yields results in agreement with experiment in many aspects~\cite{Tsai2010, Xu2012}, such as the value of the contact angle, which can be determined by macroscale equations. When the droplet has a fixed volume, the theoretical model captures a dependency of the wetting state on the geometric parameters of the surface, showing that the CB state is thermodynamically  stable for small interpillar distance and large pillar height, which is in qualitative agreement with previous findings~\cite{Lundgren2003, Shahraz2012}. When compared to experiments where the droplet evaporates, the approach describes qualitatively the dependency of the wetting state on the initial volume of the droplet: for bigger droplets, the region where CB is stable increases, as seen in Figure~\ref{transitionR}, which agrees with experiments where the droplet evaporates and changes from CB to W state below a given volume~\cite{Nosonovsky2007,Tsai2010}. Moreover, Figure~\ref{theta_B_f_vsR} shows that quantities usually  accessible experimentally like $\theta_C$ and $B$ are well described by the theoretical model when compared to the simulations initialized in the W state.  Another interesting aspect is that both theory and simulations capture the stick-slip behavior observed experimentally~\cite{Ramos2015, Enright2012}.  
Finally, we note that the MC simulations in three dimensions show a strong dependency on the initial conditions: if the initial state of the droplet is W, the agreement between theory and simulations is in general very good. When it starts at the CB state, the agreement only happens for very small pillar height, when the energy barrier is small.  These findings confirm the metastability already observed in experiments~\cite{Quere2008, McHale2005} and in 2D simulations~\cite{Shahraz2012,Shahraz2013}, as discussed in the previous section.

Both theoretical and simulational models can be modified to include other surfaces and to take into account different effects. 
Examples include:  (i) the role of reentrant surfaces; 
 (ii) the introduction of asperities at  the nanoscale on top of the pillars in the micro scale considered here (experimentally predicted to create extreme hydrophobicity~\cite{Yongjoo2009}); (iii) the  role of ordered/disordered pillar distribution on the final wetting state. 


\begin{acknowledgement}
 The authors thank Stella M. M. Ramos for stimulating us to study the phenomenon of  superhydrophobicity, for sharing her experimental data and references, and for very relevant comments on the manuscript. We thank the supercomputing laboratory at Universidade Federal da Integra\c{c}\~ao Latino-Americana (LCAD/Unila), where part of the simulations were run, for computer time. The authors also thank Tiago Vier for the artistic conception of Figure 1.
\end{acknowledgement}




\bibliography{droplet}

\begin{mcitethebibliography}{51}
\bibitem[Not()]{Note-1}
The interface sites are defined as the ones which have at least one neighbor in
  a different state than its own, among the $18$ first and second closest
  neighbors. This smaller neighborhood (compared to the Moore neighborhood) is
  used so that it is possible to obtain spherical droplets, despite the
  underlying lattice symmetries.\relax
\mciteBstWouldAddEndPunctfalse
\mciteSetBstMidEndSepPunct{\mcitedefaultmidpunct}
{}{\mcitedefaultseppunct}\relax
\EndOfBibitem
\end{mcitethebibliography}

\end{document}